\documentclass[a4paper,11pt]{article}
\usepackage{pos}

\title{Recent highlights and plans of the AWAKE experiment}

\manuallySeparateAuthors
\author*{Giovanni Zevi Della Porta}
\author{ on behalf of the AWAKE Collaboration}

\affiliation{CERN,\\
  Geneva, Switzerland}

\emailAdd{gzevi@cern.ch}

\abstract{
The Advanced Wakefield Experiment (AWAKE) is an accelerator R\&D experiment at CERN using, for the first time, a high-energy proton bunch to drive wakefields in
 plasma and accelerating electrons to the GeV energy scale. The principle of the AWAKE experiment is described. We show experimental results of the seeded 
 self-modulation process of the long 400 GeV SPS proton bunch, transforming the bunch into a train of micro-bunches and driving resonantly the wakefields in the 
 10 m long Rb plasma. We also show that externally-injected electrons can be accelerated by these wakefields to several GeV. The next steps of the AWAKE 
 experimental program are shown. Possible first applications to high-energy physics experiments, where the scheme takes advantage of the large energy stored 
 in the proton bunch to reach very high energy gain in a single plasma, are described.}

\FullConference{%
  40th International Conference on High Energy physics - ICHEP2020\\
  July 28 - August 6, 2020\\
  Prague, Czech Republic (virtual meeting)
}

\begin{document}
\maketitle

\section{Run 1 (2016-2018): setup and experimental results}

The first data taking period of the Advanced Wakefield Experiment (AWAKE) ~\cite{AwakeExperiment}
 successfully achieved two important milestones: the longitudinal self-modulation of a proton beam 
in a plasma, and the acceleration of externally-injected electrons.

\begin{figure}[!htb]
\centering
$\vcenter{\hbox{\includegraphics[width=.52\textwidth]{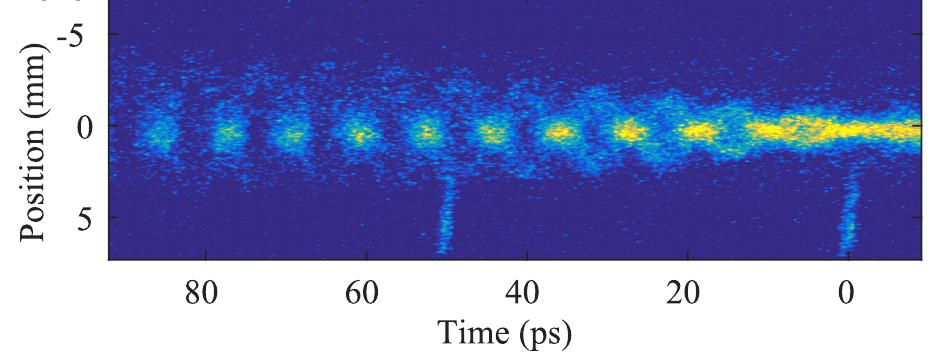} }}$
$\vcenter{\hbox{\includegraphics[width=.40\textwidth]{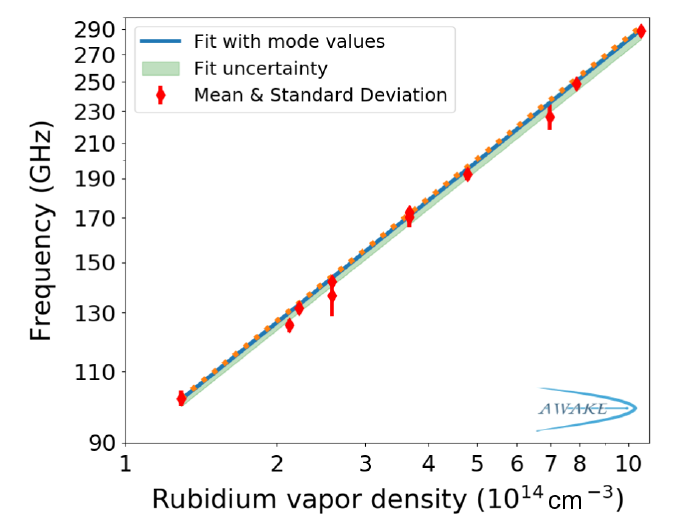} }}$
\caption{(Left) Streak camera image of self-modulated proton beam. 
(Right) Measured frequency of self-modulation as a function of plasma density~\cite{PRLSelfMod1,PRLSelfMod2,FabianSelfModStreak}.}

\label{SelfModulation}
\end{figure}

\begin{figure}[!htb]
\centering
\includegraphics[width=.49\textwidth]{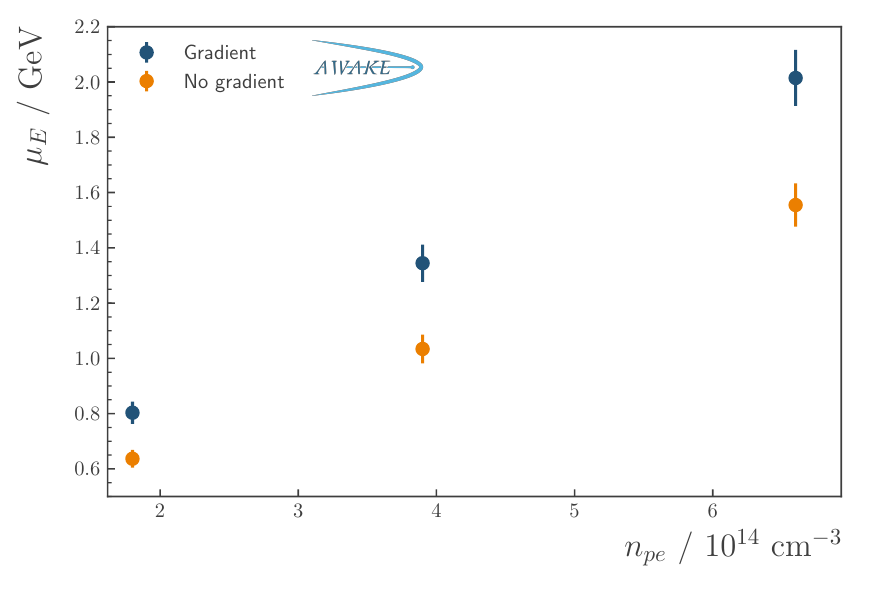}
\hspace*{.03\textwidth}
\includegraphics[width=.46\textwidth]{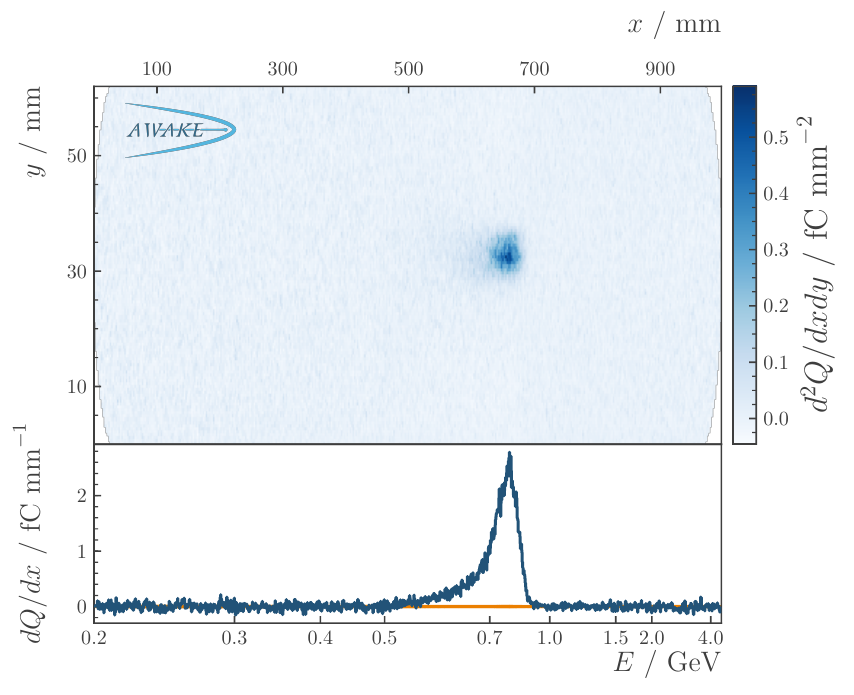} 
\caption{ (Left) Measured energy of accelerated electrons as a function of plasma density for different density gradients. 
(Right) Signal of accelerated electrons on the spectrometer scintillator for a single event~\cite{NatureAcceleration}.}

\label{Acceleration}
\end{figure}

As shown in Figure~\ref{SelfModulation}, a long proton beam is self-modulated as it interacts with the Rb plasma,  
forming micro-bunches at the characteristic frequency of the plasma.
While the frequency of the modulation can be controlled by adjusting the plasma density, the phase of the modulation is 
controlled by `seeding' the self-modulation process with a large transverse wakefield, thus creating a fully deterministic and repeatable 
process~\cite{PRLSelfMod1,PRLSelfMod2,FabianSelfModStreak}.
Electron acceleration results are summarized in Figure~\ref{Acceleration}:  a 19~MeV electron beam is injected into the self-modulated proton 
beam, and the longitudinal wakefields driven by the proton micro-bunches provide acceleration to a subset of the electrons.
Energies of up to 2~GeV are reached at the end of the 10~m long plasma, with a maximum accelerated charge of approximately 100~pC, 
corresponding to a fifth of the charge of the initial electron beam. As expected from theory and simulations, the acceleration is affected by the plasma 
density and by the density gradient along the path of the beams~\cite{NatureAcceleration}. 


\section{Run 2: goals and milestones}

Building upon the success of Run 1, an ambitious set of goals and milestones has been defined for the AWAKE Run 2 project~\cite{Run2Plan}.
The overall aim of Run 2 will be to demonstrate the scalability of the AWAKE scheme for future particle physics applications, discussed in the next section.
The goals of Run 2 are: (i) accelerate an electron beam to high energy (with a gradient of 0.5-1~GV/m), (ii) preserve electron beam quality
and (iii) demonstrate a scalable plasma source technology which can support up to and beyond 100 m of plasma.

The first milestone, in Run 2a, will be to demonstrate that the proton bunch self-modulation can be seeded by the existing electron beam. This will ensure that the 
entire proton bunch self-modulates, as opposed to only the second half of the bunch, as in the current configuration~\cite{ElectronSeeding}. 
The second milestone, in Run 2b, will be to demonstrate that the self-modulation process can be stabilized when the longitudinal wakefields 
reach their maximum amplitude, ensuring that the wakefields can be sustained for long timescales. This will require a sharp change in the plasma 
density, which will be possible in a dedicated plasma cell designed specifically for self-modulation~\cite{PathToAwakeDensityStep}.
Run 2c will require an additional plasma cell, to be installed downstream of the self-modulation cell, and a new electron beam, which will be injected 
between the two cells. This configuration will avoid interference between the self-modulation and acceleration processes, and it will use two  
known effects, beam loading the longitudinal wakefield and full blowout of plasma electrons, to preserve the quality of the electron beam~\cite{Veronica}.

The last goal, a scalable plasma source, can be achieved independently of the other two. 
The laser-ionization scheme currently used to generate the plasma cannot support long plasma cells, so two alternative schemes are currently being 
studied: one based on a low-frequency electromagnetic wave generated by RF antennas (Helicon)~\cite{Helicon} and one based on a high-current arc 
in the plasma (Discharge)~\cite{Discharge}

\section{Beyond Run 2: applications to high-energy particle physics}

A roadmap for particle physics applications of the AWAKE acceleration scheme has been laid out in a set of documents prepared for the 
European Particle Physics Strategy 
Update~\cite{EuropeanPathToApplications,EuropeanApplications,EuropeanAwakePlusPlus}.
The roadmap takes into account our current knowledge, and proceeds from the more realistic goals to those that will be harder to achieve.

\begin{figure}[ht]
\centering
$\vcenter{\hbox{\includegraphics[width=.32\textwidth]{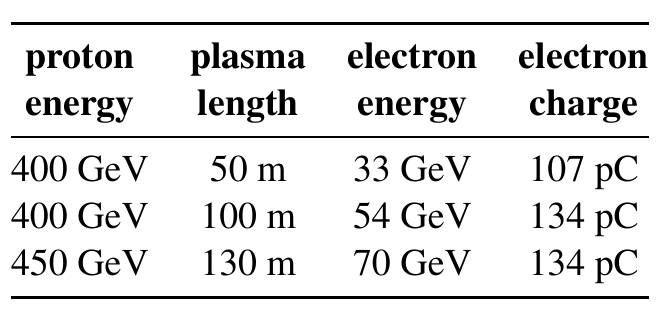} }}$
\hspace*{.03\textwidth}
$\vcenter{\hbox{\includegraphics[width=.62\textwidth]{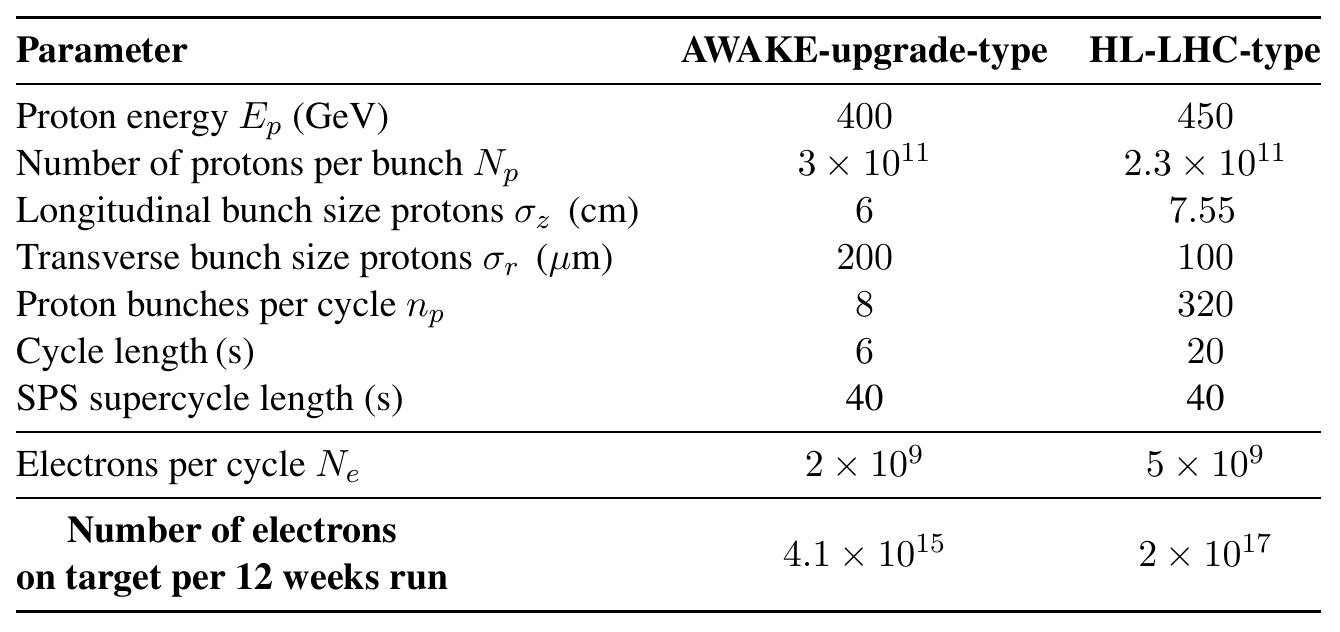} }}$
\caption{(Left) Electron energies achievable with different plasma lengths. 
(Right) Potential achievable number of electrons on target for different AWAKE-scheme fixed target configurations. 
For reference, the NA64 experiment at CERN expects $3\times10^{12}$ electrons during its entire lifetime~\cite{EuropeanAwakePlusPlus}.}

\label{FixedTargetTables}
\end{figure}

\begin{figure}[ht]
\centering
\includegraphics[width=.46\textwidth]{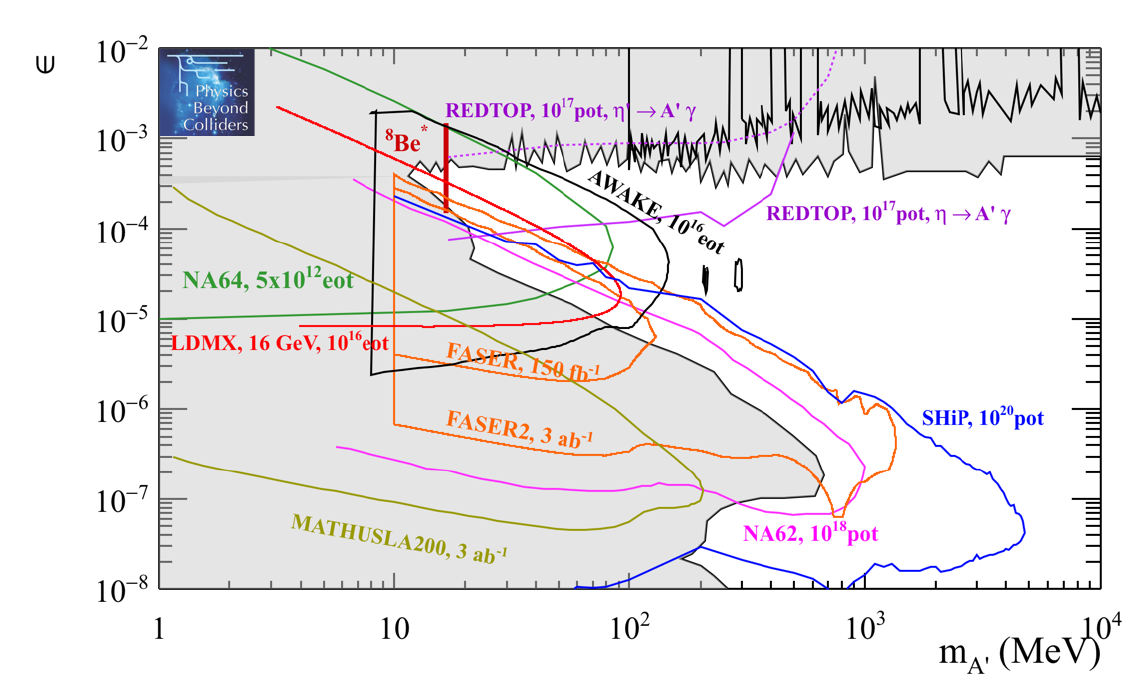}
\hspace*{.03\textwidth}
\includegraphics[width=.475\textwidth]{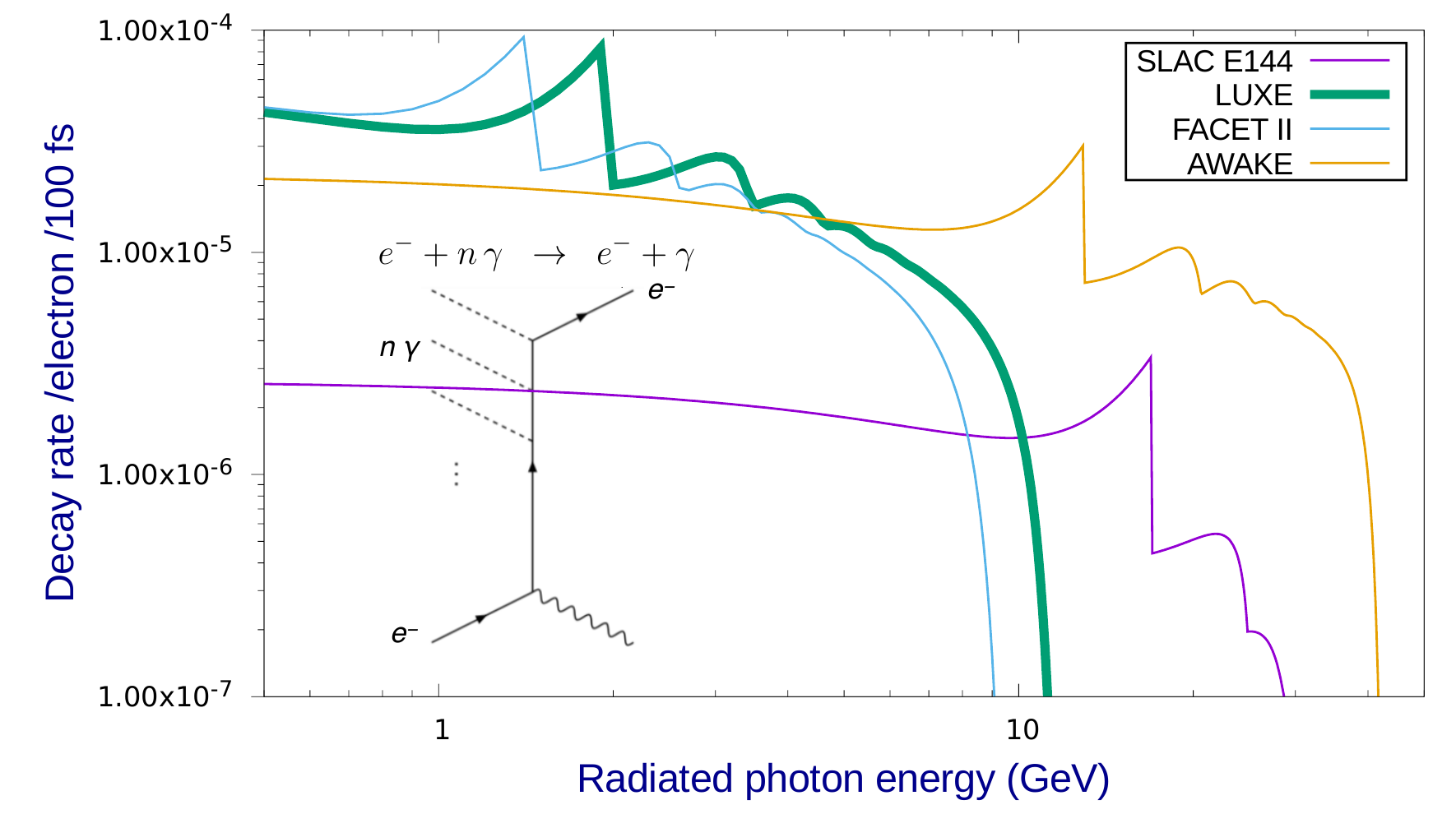}
\caption{
(Left) Dark photon sensitivity, as a function of coupling and mass, for an AWAKE-scheme fixed target experiment (black line), 
compared to existing limits (grey) and ongoing/planned experiments (colored lines).
(Right) Strong-field QED sensitivity, in terms of electron/laser interaction rate as a function of radiated photon energy, for an AWAKE-scheme experiment, 
compared to other experiments~\cite{PhysicsBeyondColliders,StringFieldQED, EuropeanApplications}.}

\label{FixedTarget}
\end{figure}

Fixed target experiments would be the first application of AWAKE. By extending the acceleration plasma cell from 10~m to O(100)~m, energies
of the order of 50~GeV would be achievable. While the transverse beam quality would be too low for a collider, it would be sufficient for 
a fixed target experiment, and the electron flux would be several orders of magnitude higher than existing experiments at this energy, 
as shown in the tables of Figure~\ref{FixedTargetTables}.
Such electron beams could significantly extend the reach of searches for rare particles such as dark photons, as well as probe
strong-field QED electron/laser interactions, as shown in Figure~\ref{FixedTarget}.

\begin{figure}[ht]
\centering
\includegraphics[width=.85\textwidth]{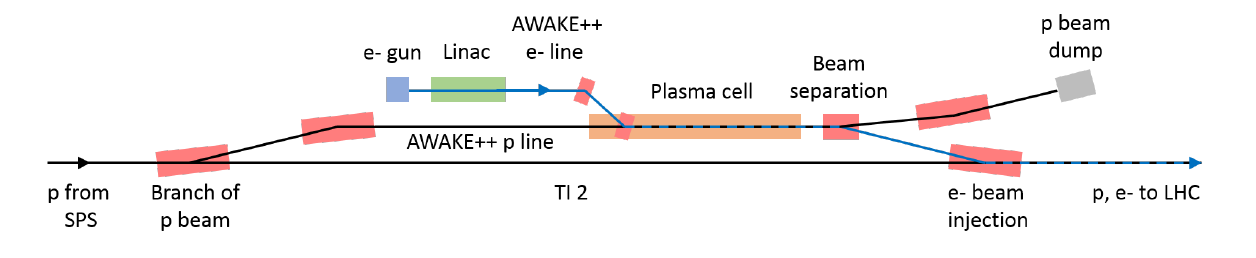}
\includegraphics[width=.48\textwidth]{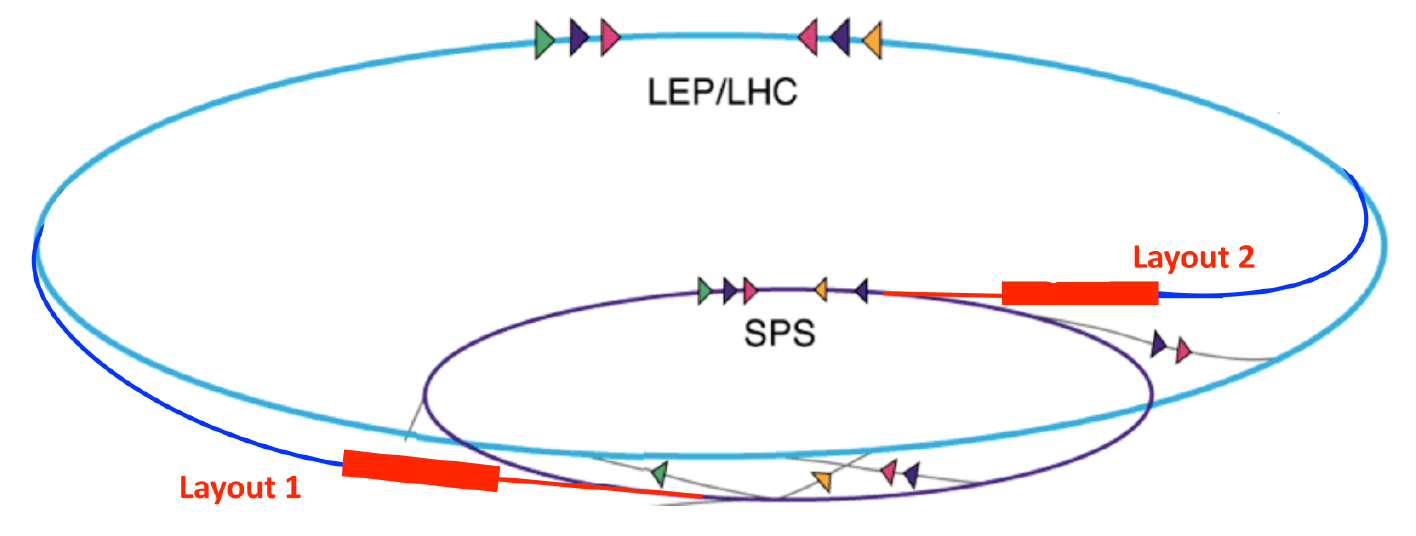}
\hspace*{.07\textwidth}
\includegraphics[width=.35\textwidth]{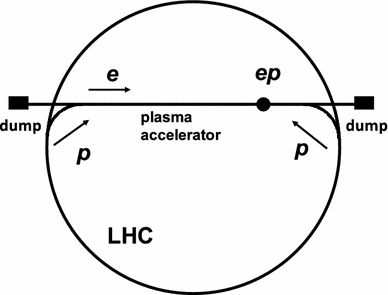}
\caption{
(Top) Schematic layout of an AWAKE-scheme electron accelerator for an electron/proton or electron/ion collider.
Schematic layouts of the Plasma Electron Proton/Ion Collider (PEPIC), reaching a center-of-mass energy
of 1.3~TeV (left), and the Very High Energy electron-Proton collider (VHEeP), reaching 9~TeV (right)~\cite{PEPIC,VHEeP}.}
\label{epSchematics}
\end{figure}

\begin{figure}[ht]
\centering
\includegraphics[width=.45\textwidth]{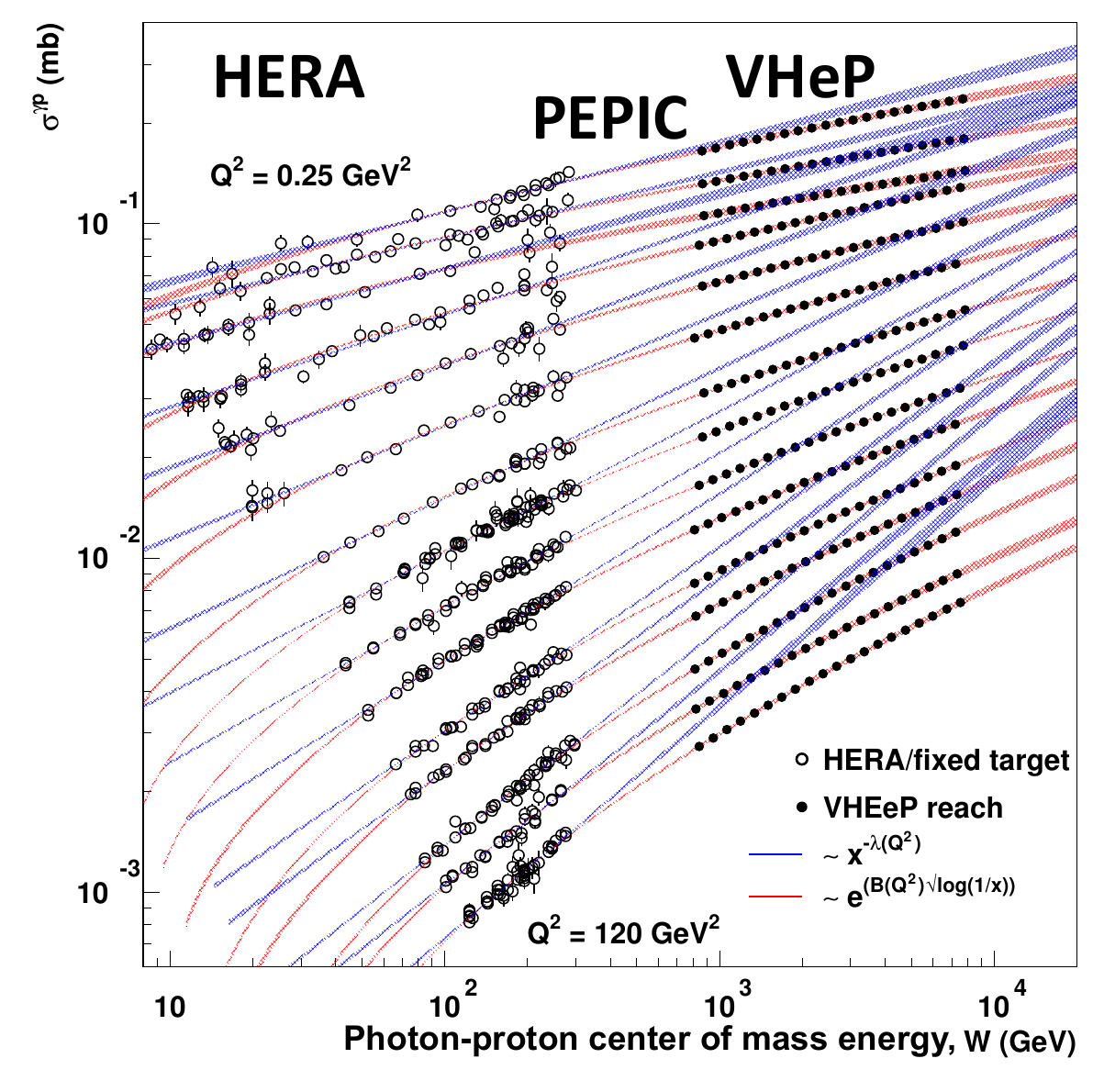}
\hspace*{.03\textwidth}
\includegraphics[width=.455\textwidth]{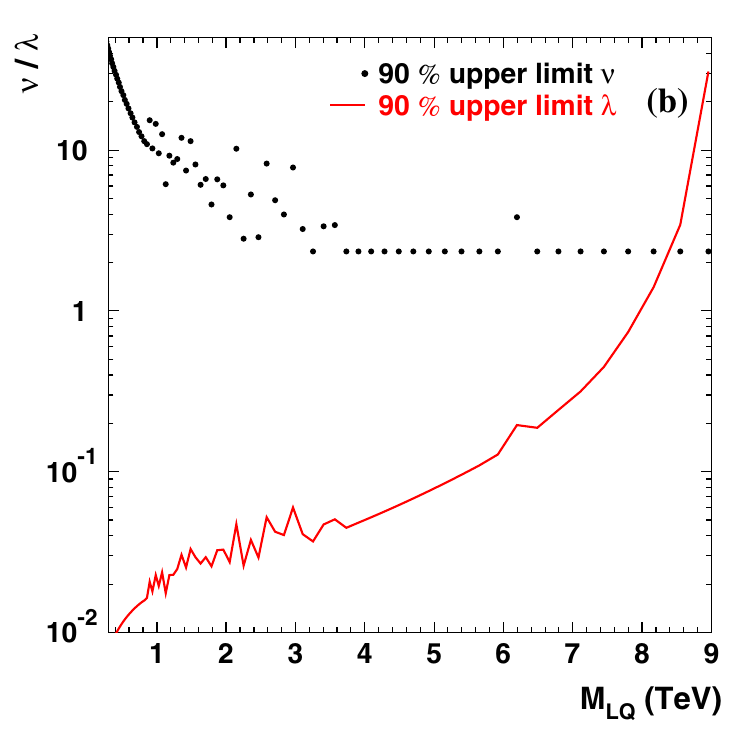}
\caption{
Physics reach of the Very High Energy electron-Proton collider (VHEeP).
(Right) Measurement of photon-proton cross sections versus the center of mass energy, extending the reach of HERA.
(Left) Upper limits on the number of events, $\nu$, and leptoquark coupling parameter, $\lambda$, versus mass of the leptoquark, 
$\mathrm{M}_{\mathrm{LQ}}$~\cite{VHEeP}.}
\label{VHEeP_sigma_gp}
\end{figure}

As long as positron acceleration in wakefields remains a challenge, planning for an electron-positron machine is not realistic.
However, the AWAKE scheme allows transformation of an existing hadron collider into an electron-proton or electron-ion collider,
by using one of the proton beams to accelerate electrons to high energy, and colliding these electrons into the remaining 
proton/ion beam, as shown in Figure~\ref{epSchematics}.
Such  electron-proton (or electron-ion) colliders would open a new energy regime for Deep Inelastic Scattering, 
significantly improving our current understanding of the proton/ion structure, and they would also extend the reach of direct searches for 
new particles such as leptoquarks, as shown in Figure~\ref{VHEeP_sigma_gp}.

To summarize, in Run 1 we have demonstrated that electrons can be accelerated in a plasma using a long proton bunch, taking 
advantage of the self-modulation process. 
In Run 2 we aim to control the beam quality and to demonstrate that we can build O(100) m long plasma cells. 
Assuming Run 2 will be successful, we have started considering particle physics experiments which could be built
using the AWAKE acceleration scheme. 
We explored a few examples for fixed target and electron-proton collisions, but of course there are many more opportunities to consider.

\bibliographystyle{JHEP}
\bibliography{zeviAWAKE}

\end{document}